\documentclass[preprint,amsmath]{revtex4-1}

\usepackage{blindtext}
\usepackage{epsfig}
\usepackage{amsmath}
\usepackage{amssymb}
\usepackage{color}
\usepackage[normalem]{ulem}
\usepackage{caption}
\usepackage{subcaption}
\usepackage{mathtools}
\usepackage{color}
\usepackage[T1]{fontenc}
\usepackage{epstopdf}
\usepackage{array}
\usepackage{wasysym}
\usepackage{mathrsfs}
\usepackage{graphicx}
\usepackage{geometry}

\begin{document}

\title{The case of the biased quenched trap model in two dimensions with diverging mean dwell times.}
\author{Dan Shafir}
\email{dansh5d@gmail.com}
\author{Stanislav Burov}
\email{stasbur@gmail.com}
\affiliation{Physics Department, Bar-Ilan University, Ramat Gan 5290002,
Israel}
\date{\today}

\begin{abstract}
We investigate the biased quenched trap model on top of a two-dimensional lattice in the case of diverging expected dwell times.
By utilizing the double-subordination approach and calculating the return probability in $2$d, we explicitly obtain the disorder averaged probability density function of the particle's position as a function of time  (for any given bias) in the limit of large times ($t \rightarrow \infty$). 
The first and second moments are calculated, and a formula for a general $\mu$-th moment is found. 
The behavior of the first moment, i.e. $\langle x(t)\rangle$, presents non-linear response both in time and in the applied external force $F_0$. While the non-linearity in time occurs for any measurement time $t$, the non-linearity in $F_0$ is expected only when $t\gtrsim \big(F_0 \left| \ln (F_0)\right|\big)^{-2 / \alpha}$ where $\alpha=T/T_g$, for temperatures $T<T_g$. 
We support our analytic results by comparison to numerical simulations.
\end{abstract}

\maketitle

\section{Introduction} \label{se: Introduction}
Properties of transport in a disordered medium where the motion is dictated by diverging expected waiting (or dwell) times have been the subject of rigorous research in many fields \cite{RevModPhys.53.175, PhysRep.195.127, PhysRep.339.1}. 
Such systems exhibit anomalous slow diffusion, termed subdiffusion, i.e the second moment grows with time as $\langle x^2(t) \rangle  \sim t^{\alpha}$ where $0< \alpha <1$. 
This type of diffusion describes motion in complex disordered systems such as living cells \cite{PhysToday.65.8.29, PNAS.110.4911}, blinking quantum dots \cite{PhysToday.62.2.34}, molecular-motor transport on a filament network \cite{PhysRevX.6.011037} and photocurrents in amorphous materials \cite{PhysRevB.12.2455}. 
It was also shown that anomalous diffusion could be used to describe the aging phenomenon in glasses \cite{JPhysI.2.1705, JPhysA.29.3847, PhysRevLett.84.5403, PhysRevB.64.104417, RevModPhys.83.587}.

The basic model to describe transport in disordered material is the random walk (RW) on a lattice. One of the most popular generalizations of the RW is the continuous-time random walk (CTRW) that was developed in the works of  Montroll, Scher, and others \cite{PhysRevB.12.2455, JStatPhys.10.421, PhysRevLet.44.55, PhysToday.44.1.26} as a description of transport in amorphous materials.
CTRW describes the motion of a traced particle affected by traps. The particle can be trapped in various locations of the media, where it waits for some random time before continuing its motion. 
In CTRW, the particle hops from site to site on top of a lattice. Between the hops, a random dwell time is generated.   
When the expected dwell times diverge, the behavior is non-ergodic \cite{PhysRevLett.94.240602, PhysRevLett.101.058101} and  subdiffusive.
The dwell times between hops for a CTRW process on top of a lattice are assumed to be independent, identically distributed (i.i.d) random variables. This situation of uncorrelated disorder is termed annealed disorder, and it allows mathematical treatment utilizing the renewal theory.
The disorder when the dwell times are position-dependent is termed quenched disorder.
In the quenched trap model (QTM), the dwell time at each lattice site stays fixed for the whole time. This means that if the particle revisits the same site, it stays there for the same dwell time as during the previous visitation.
For quenched disorder, the renewal assumption is lost, at least below a critical dimension, since revisiting a site invokes previously experienced dwell time. 
Nevertheless, scaling arguments and the renormalization group approach \cite{PhysRep.195.127, JPhysA.18.L531, PhysRevE.68.036114} suggests that for dimension $d > 2$, QTM behaves like CTRW, which in some sense represents a mean-field approximation for the QTM. 
In \cite{ProbTheo.134.1, Ann.Probab.35.2356,Ann.Probab.43.2405, StochProc.125.1032} a similar conclusion was obtained by using a rigorous mathematical description of QTM on a regular lattice  and a more general approach of randomly trapped random walks. 
The correspondence between QTM and CTRW is exact when $d \rightarrow \infty$. 
The logic behind this claim stems from the fact that in high  dimensions ($d>2$), the particle is non-recurrent  and rarely visits the same site more than once. So by increasing the dimensionality, we reduce our correlations and approach the annealed case of CTRW. 
Below the critical dimension of $2$ (the particle is recurrent), correlations imposed by identical dwell times become apparent and cannot be ignored. 
Thus, when considering an RW in a  system with quenched disorder, correlations in space (due to multiple visits to the same site) make the RW much harder to handle mathematically.

While an abundance of research on the  subject  at  hand  is present, a direct mapping between QTM and CTRW has been lacking until recently.
Recent works \cite{PhysRevE.96.050103, JStatMech2020.073207} have shown that a temporal transformation can take us from CTRW to the equivalent time in QTM, in the case of transient RW where the probability of return to a previously visited site ($Q_0$) is less than 1. 
Specifically, this temporal transformation is valid for systems with dimension $d> 2$ or any biased walk (i.e. external force is applied)  with quenched trapping disorder~\cite{PhysRevLett.110.067801, PhysRevLett.106.090602, PhysRevLett.111.260601, PhysRevLett.116.138301}. 
This mapping between CTRW and QTM was the result of introducing a new local time parameter ($S_{\alpha}$), which allowed to calculate the QTM's spatial  probability density function (PDF), averaged over disorder, for the biased one-dimensional case \cite{JStatMech2020.073207}. 
The QTM's PDF of the particle's position on top of a two-dimensional lattice has yet to be found and compared to numerical results. Hence the purpose of this paper.

While the one-dimensional and two-dimensional situations share similar features, like the infinite average number of steps it takes the particle to return to the origin, these cases are qualitatively different.
To better understand the difficulties caused by working in two dimensions, we mention a problem discussed by Weiss \cite{WeissBook}. Consider a set of $k$ $(k>0)$ independent non-interacting walkers. Each performing unbiased RW, and all are starting from the same starting position simultaneously. Assuming that the RW is recurrent, we consider the average return time of the earliest walker to return to the origin. 
The question is, how large $k$ has to be to ensure that the first one to return to the origin  has a finite average return time? In $1d$, the answer is $k>2$, and in fact, the first  $(k-2)$ returning RWs will have a finite average return time.
For the two-dimensional case, for any $k>0$, none of the walkers will have a finite average return time. 
This result emphasizes the difficulty of working in two dimensions.
Therefore the properties of quenched disorder in two-dimensional QTM need to be handled with much care.

This paper is organized in the following manner. In Sec. (\ref{se: Theoretical background}), we provide a brief review of the theoretical background and a formula for the $\mu$-th moment of the position, together with the particle's diffusion front. 
In Sec. (\ref{se: Response to a bias}), we explicitly evaluate the first and second moments which exhibit anomalous diffusion and non-linear response for the case of the rectangular lattice. In the appendix (\ref{se: Appendix}), we provide examples for the evaluation of the moments on other lattice types, such as the oblique and hexagonal lattices.
Sec. (\ref{se: Equivalency of the QTM to CTRW}) presents the equivalency between the QTM and CTRW. A temporal mapping between the two is given, and a comparison between QTM in one and two dimensions is performed.
In Sec. (\ref{se: Simulations}) we discuss the simulations and validate our theoretical results.
A formula for the minimal measurement time required for the theory and simulations to converge is achieved. 
Sec. (\ref{se: Conclusions}) presents our conclusions and relates our results to other works done in the field.

\section{Theoretical background} \label{se: Theoretical background}
For the CTRW model with diverging mean dwell-times, the process is decomposed into ordinary spatial  Brownian motion and a temporal L\' {e}vy process, an approach called subordination \cite{Chaos.7.753, PhysRevE.63.046118, JApplProbab.41.623, Chaos.15.026103, PhysRevE.72.061103}. 
In QTM, the disorder is quenched, the dwell times are correlated, and L\' {e} vy's limiting theorem cannot be applied since the dwell times are not i.i.d. To solve this problem, we follow recent works \cite{PhysRevE.96.050103, JStatMech2020.073207, PhysRevE.86.041137, PhysRevLett.106.140602} and define a new local time parameter together with an approach termed double subordination to solve a biased QTM on a 2d lattice with diverging mean dwell times. As was mentioned in Ref. \cite{PhysRevE.96.050103} when QTM is biased, the return probability is $<1$, the quenched nature of QTM is less apparent (fewer re-visitations of lattice sites), and one can assert those correlations imposed by the quenched dwell times can be effectively renormalized into uncorrelated times, i.e., a CTRW description. Thus a mapping between QTM and CTRW can be achieved. In this section, we summarize the results received in a recent paper by one of the authors \cite{JStatMech2020.073207} and expand upon them in the following sections to obtain the moments for the case of the transient (nonrecurring) QTM affected by a bias in two dimensions. We refer the interested reader to the original paper \cite{JStatMech2020.073207} for rigorous derivations.

Our first step is to define the problem and present the local time used in the case of the quenched disorder.
Before doing that, to better understand the thought process, we will show why the local time in the case of CTRW (annealed disorder) is simply the number of jumps $N$. 
For a given number of waiting times $N$, the total measurement time $t$  is 
\begin{equation}
    t=\sum_{i=1}^{N} \tau_i
\end{equation}
where $\tau_i$ are the local dwell times, i.e., i.i.d random variables distributed according to the PDF $\omega (\tau_i) = A\tau_i^{-(1+\alpha)}/|\Gamma(-\alpha)|$ where $0<\alpha<1$, $A>0$ and $\Gamma()$ is the Gamma function. 
For these values of $\alpha$, the average of each $
\tau_i$ is diverging.
It is known \cite{PhysRep.195.127} that in the large $N$ limit
\begin{align} \label{eq:t relation}
\text{PDF}\left(\frac{t}{N^{1/\alpha}}\right) = l_{\alpha,A,1}\left(\frac{t}{N^{1/\alpha}}\right)
\end{align}
where $l_{\alpha,A,1}()$ is the one sided L\'{e}vy distribution~\cite{PhysRevLett.106.140602} which is defined by its Laplace transform, $\int_0^\infty e^{-Z u} l_{\alpha,A,1}(Z)\,dZ= e^{-Au^{\alpha}}$.
We now have the probability of observing $t$ for a given number of steps $N$. Then, the probability of observing $N$ steps during time $t$ can be found \cite{PhysRep.195.127, JStatMech2020.073207} by changing variables from $Z=t/N^{1/ \alpha}$ to $N=(t/Z)^{\alpha}$ with the result
\begin{align} \label{eq: Pt(N)}
\mathscr{P}_{t}(N)=\frac{t}{\alpha} N^{-1-1 / \alpha} l_{\alpha, A, 1}\left(\frac{t}{N^{1 / \alpha}}\right) \quad(t \rightarrow \infty).
\end{align}
We  condition on the different outcomes of $N$ (in the limit of $t \rightarrow \infty$) and obtain for the PDF to observe a particle in position $\mathbf{r}$ at time $t$,
\begin{equation} \label{eq: PDF CTRW}
\langle P(\mathbf{r}, t)\rangle_{CTRW}=\sum_{N} W_{N}(\mathbf{r}) \mathscr{P}_{t}(N),
\end{equation}
where $\langle \dots \rangle$ is an average upon disorder (many realizations) and $W_N (\mathbf{r})$ is the PDF of regular RW irrespective of the various dwell times, i.e can be approximated by a Gaussian in the large $N$ limit (by using the central limit theorem).
Eq. (\ref{eq: PDF CTRW}) shows that since in the PDF of the regular RW, i.e., $W_N (\mathbf{r})$, the time is replaced by $N$, hence  $N$  is the natural choice for the local time in the case of CTRW. 
This procedure of separating CTRW into spatial and temporal processes is termed subordination.

Now we return to our main subject of the QTM and apply a similar approach by finding the appropriate local time. During the measurement time $t$, the particle has visited a certain amount of lattice points and stayed exactly $\tau _{\mathbf{r}}$ at each lattice point $\mathbf{r}$. The quenched dwell times {$\tau _{\mathbf{r}}$} are real, positive, and independently distributed random variables with 
\begin{align}
\psi\left(\tau_{\mathbf{r}}\right) \sim \tau_{\mathbf{r}}^{-(1+\alpha)} A /|\Gamma(-\alpha)| \quad\left(\tau_{\mathbf{r}} \rightarrow \infty\right)
\end{align}
as the PDF ($A > 0$ and $\Gamma()$ is the Gamma function). The value of the exponent $\alpha$ is bounded to $0 < \alpha < 1$. For these values of $\alpha$, the average dwell times diverge, i.e, $\langle \tau_{\mathbf{r}} \rangle=\int_{\tau_{min}}^{\infty} \tau_{\mathbf{r}} \psi(\tau_{\mathbf{r}}) d\tau_{\mathbf{r}}=\infty$ and the model results in anomalous sub-diffusion and aging \cite{JPhysI.2.1705}. 
The physical representation of QTM assumes a thermally activated particle that is jumping between various energetic traps.
When the particle is in a trap located at $\mathbf{r}$, the mean escape time $\tau_{\mathbf{r}}$ is given by Arrhenius law $\tau_{\mathbf{r}} \sim \exp (E_{\mathbf{r}}/T)$, where $E_{\mathbf{r}}$ is the depth of the trap and $T$ is the temperature. 
When the PDF of $E_{\mathbf{r}}$ is exponential, i.e., $\frac{1}{T_g} \exp (-E_{\mathbf{r}}/T_g)$, the mean escape time achieves a PDF in the from of $\psi (\tau_{\mathbf{r}})$ where $\alpha=T/T_g$. 
For low temperatures ($T<T_g$), one observes glassy behavior (aging and non-self averaging) in the system \cite{PhysRevE.67.026128}.

The measurement time t is provided by
\begin{align}
t=\sum_{i=0}^{N} \tau_i=\sum_{\mathbf{r}} n_{\mathbf{r}} \tau_{\mathbf{r}} 
\end{align}
where $n_{\mathbf{r}}$ is the number of times the particle visited site ${\mathbf{r}}$ during $t$ and the summation is over all the lattice points.
In the case of  QTM, the quantity
\begin{equation}
S_{\alpha}=\sum_{\mathbf{r}}\left(n_{\mathbf{r}}(t)\right)^{\alpha}
\end{equation}
can serve as the local time in the same way that the number of jumps $N$ (performed during the time t) is the local time for CTRW. 
This is true based on the fact that in the limit of $t\rightarrow \infty$
\begin{align} \label{eq:t quenched relation}
\text{PDF}\left(\frac{t}{S_{\alpha}^{1/\alpha}}\right) = l_{\alpha,A,1}\left(\frac{t}{S_{\alpha}^{1/\alpha}}\right)
\end{align}
which was proven in \cite{PhysRevE.96.050103, JStatMech2020.073207, PhysRevE.86.041137}.
Eq. (\ref{eq:t quenched relation}) is the QTM's analog of the CTRW (Eq. (\ref{eq:t relation})) where we have switched the local time $N$ with $S_{\alpha}$. Note that $S_{\alpha}$ is a spatial variable that depends solely on various positions of the particle and not the time spent at those sites. 
For $\alpha = 1$ $S_{\alpha}$ is the total number of steps performed, and for $\alpha=0$ is the total number of distinct sites visited during $t$. 

The probability of observing a specific $S_{\alpha}$ for a given measurement time $t$ (i.e., $\mathcal{P}_{t}\left(S_{\alpha}\right)$) is now obtained from Eq.  in a similar fashion to how Eq. (\ref{eq: Pt(N)}) was obtained from the relation in Eq. (\ref{eq:t relation})~\cite{JStatMech2020.073207}:
\begin{equation} \label{eq: pr of S_alpha}
\mathcal{P}_{t}\left(S_{\alpha}\right) \sim \frac{t}{\alpha} S_{\alpha}^{-1 / \alpha-1} l_{\alpha, A, 1}\left(\frac{t}{S_{\alpha}^{1 / \alpha}}\right) \quad(t \rightarrow \infty).
\end{equation}
The PDF $P({\mathbf{r}}, t)$ to find the particle at position ${\mathbf{r}}$ after measurement time $t$ is calculated by conditioning on all the possible $S_{\alpha}$ that can occur during the process. 
One needs to sum over all the possible $P_{S_{\alpha}}(\mathbf{r})$ (i.e. the PDF to observe the particle at $\mathbf{r}$ for a given $S_\alpha$) multiplied by the appropriate probability to observe such $S_{\alpha}$ at time $t$, for a given disorder. After averaging over disorder, the PDF takes the form 
\begin{equation} \label{eq: Sum1}
\langle P(\mathbf{r}, t)\rangle=\sum_{S_{\alpha}} P_{S_{\alpha}}(\mathbf{r}) \mathcal{P}_{t}\left(S_{\alpha}\right)
\end{equation}
and due to Eq. (\ref{eq: pr of S_alpha}), in the $t \rightarrow \infty$ limit we obtain 
\begin{equation}
\langle P(\mathbf{r}, t)\rangle \sim \int_{0}^{\infty} P_{S \alpha}(\mathbf{r}) \frac{t}{\alpha} S_{\alpha}^{-1 / \alpha-1} l_{\alpha, A, 1}\left(\frac{t}{S_{\alpha}^{1 / \alpha}}\right) d S_{\alpha}
\end{equation}

The PDF $P(\textbf{r}, t)$ depends on $P_{S_{\alpha}}(\textbf{r})$. The form of $P_{S_{ \alpha}}(\mathbf{r})$ is obtained by using again the subordination approach where the local time $S_{\alpha}$ is subordinated to $N$, number of jumps performed, and the spatial process is provided by $W_N (\mathbf{r})$—the PDF of regular RW, i.e.
\begin{equation} \label{eq: Sum2}
P_{S_{\alpha}}(\mathbf{r})=\sum_{N=0}^{\infty} W_{N}(\mathbf{r}) \mathcal{G}_{S_{\alpha}}(N, \mathbf{r})
\end{equation}
where $\mathcal{G}_{S_{\alpha}}(N, \mathbf{r})$ is the probability to perform $N$ steps before reaching $\textbf{r}$ provided that the value of $S_{\alpha}$ is known.

Now we need to find an explicit expression for $P_{S_{\alpha}}(\mathbf{r})$ to establish a simplified representation of the positional PDF, i.e. $P(\textbf{r}, t)$. From \cite{PhysRevE.96.050103, JStatMech2020.073207} we know that for the case when the spatial process is transient and the probability of eventually returning to the origin $Q_0$, is less than 1, in the large $N$ limit, the local time $S_{\alpha}$ and the number of jumps $N$ obey  linear dependence 
\begin{equation} \label{eq: QTM local time linear connection}
\big\langle S_{\alpha} (N) \big\rangle=\Lambda N \; \; \; (N \rightarrow \infty)
\end{equation}
where
\begin{equation} \label{eq: Lambda theory}
\Lambda=\frac{\left(1-Q_{0}\right)^{2}}{Q_{0}} L i_{-\alpha}\left(Q_{0}\right)
\end{equation}
and $L i_{a}(b)=\sum_{j=1}^{\infty} b^{j} / j^{a}$ is the Polylogarithm function.
\begin{figure}[t]
\centering
\includegraphics[width=0.5\textwidth]{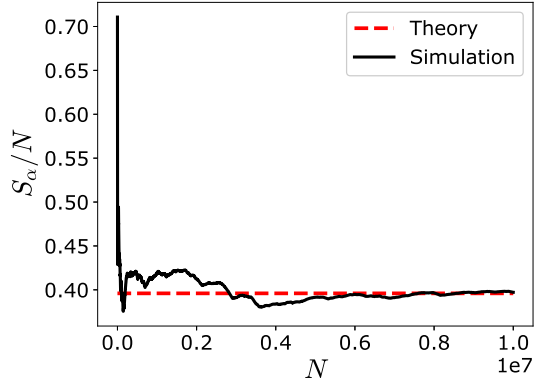}
\caption{An example for the convergence of $S_{\alpha}(N)$ to $\Lambda N$ for the case of a two-dimensional RW on a square lattice (with unit spacing),$\alpha=0.5$. 
The external force is $\vec{F}=0.001(\hat{x}+\hat{y})$
which means, following Eq. (\ref{eq: Q_0 final}), that the return probability $Q_0 = 0.812979$. 
The thick line represents the simulation while the dashed line is the theoretical limit provided by Eq. (\ref{eq: Lambda theory}).}
\label{fig:1}
\end{figure}
Moreover, it was proven in detail in Ref. \cite{JStatMech2020.073207} that for any value of $\textbf{r}$
\begin{equation} \label{eq: limit G}
\mathcal{G}_{S_{\alpha}}(N, \mathbf{r}) \underset{S_{\alpha} \rightarrow \infty} \longrightarrow \delta\left(S_{\alpha}-\Lambda N\right).
\end{equation}
This result stems from the fact that $\langle S_{\alpha} \rangle \rightarrow \Lambda N$ while also $ \text{Var}(\langle S_{\alpha}\rangle) \rightarrow 0 $.
An example for the convergence of $S_{\alpha}/N$ to $\Lambda$ can be observed in Fig. \ref{fig:1}.
By plugging Eq. (\ref{eq: Sum2}) into Eq. (\ref{eq: Sum1}) we perform a {\bf double subordination} that prescribes the disorder averaged PDF $\langle P(\textbf{r}, t) \rangle$ the form
\begin{equation} \label{eq:doublesubform}
\langle P(\mathbf{r}, t)\rangle=\sum_{S_{\alpha}} \sum_{N=0}^{\infty} W_{N}(\mathbf{r}) \mathcal{G}_{S_{\alpha}}(N, \mathbf{r}) \mathcal{P}_{t}\left(S_{\alpha}\right).
\end{equation}
Note that by taking the limit, $t \rightarrow \infty$, only large $S_{\alpha}$ need to be considered. In addition, the regular practice of the subordination technique is to replace the sums in~\eqref{eq:doublesubform} by integrals~\cite{PhysRep.195.127}.  
Therefore, by substituting the value for $\mathcal{G}_{S_{\alpha}}(N, \mathbf{r})$ from Eq. (\ref{eq: limit G}), 
the form of $\mathcal{P}_{t}\left(S_{\alpha}\right)$ in Eq. (\ref{eq: pr of S_alpha}) finally gives us the expression for the positional PDF
\begin{equation} \label{eq: probability final}
\langle P(\mathbf{r}, t)\rangle \sim \int_{0}^{\infty} W_{N}(\mathbf{r}) \frac{t / \Lambda^{1 / \alpha}}{\alpha} N^{-1 / \alpha-1} l_{\alpha, A, 1}\left(\frac{t / \Lambda^{1 / \alpha}}{N^{1 / \alpha}}\right) \mathrm{d} N \quad t \rightarrow \infty
\end{equation}
By acquiring the positional PDF we can now calculate any moment $\mu$ of the QTM, i.e., $\langle \mathbf{|r|}^{\mu} \rangle$ by its definition,
\begin{equation} \label{eq: general QTM moments}
\left\langle|\mathbf{r}|^{\mu}\right\rangle_{\text{QTM}}=\int|\mathbf{r}|^{\mu} \langle P(\mathbf{r}, t)\rangle \mathrm{d} \mathbf{r}.
\end{equation}
Since the positional PDF is a function of the displacement ($\mathbf{r}$) and the number of steps $N$, we can evaluate the displacement dependent part of the function separately. This part happens to be the moments of regular RW,
\begin{equation} \label{eq: moments spatial process}
\langle |\mathbf{r}|^{\mu} \rangle_{\text{RW}}=\int|\mathbf{r}|^{\mu} W_{N}(\mathbf{r}) \mathrm{d} \mathbf{r}=\sum_{i=1}^{\mu} B_{\mu}^{(i)} N^{\gamma_{\mu}^{(i)}}.
\end{equation}
The right-hand side of Eq. (\ref{eq: moments spatial process}) is the general solution of the $\mu$-th moment with the constants $B_{\mu}$ and $\gamma_{\mu}$ which depend on the bias (force) and the lattice type (including dimension). In the next section, we will explicitly evaluate these constants for square and rectangular lattices. In the appendix, we consider other examples - the oblique and hexagonal lattice types. Eq. (\ref{eq: moments spatial process}) can be evaluated using the central limit theorem. In the limit of large $N$ the positional PDF $W_{N} (\mathbf{r})$ of a regular RW after $N$ steps attain the Gaussian form 
\begin{equation}
W_{N}(\mathbf{r})=\frac{1}{\sqrt{2 \pi N {\text{Var}(l)}^{2}}} \exp \left[-\frac{(\mathbf{r}-\langle l \rangle N)^{2}}{2 N {\text{Var}(l)}^{2}}\right]
\end{equation}
with single-step displacement $\langle l \rangle$ and variance $\text{Var}(l)$. Now all the moments of regular RW can be easily computed. Then according to Eq. (\ref{eq: probability final}), the $\mu$th moment for the QTM is provided by substituting Eq. (\ref{eq: moments spatial process}) into Eq. (\ref{eq: general QTM moments}) with the general result
\begin{equation}
\left\langle|\mathbf{r}|^{\mu}\right\rangle_{\text{QTM}}=\sum_{i=1}^{\mu} \int_{0}^{\infty}\left(B_{\mu}^{(i)} t / \Lambda^{1 / \alpha} \alpha\right) N^{\gamma_{\mu}^{(i)}-1-1 / \alpha} l_{\alpha, A, 1}\left(t /(\Lambda N)^{1 / \alpha}\right) \mathrm{d} N
\end{equation}
and utilizing the identity \cite{PhysRevE.63.046118}
\begin{equation}
\int_{0}^{\infty} y^{q} l_{\alpha, A, 1}(y) \mathrm{d} y=A^{q / \alpha} \Gamma[1-q / \alpha] / \Gamma[1-q], \qquad (q / \alpha<1)
\end{equation}
the expression for the moments of $\mathbf{r}$ takes the final form
\begin{equation} \label{eq: final moments}
\left\langle|\mathbf{r}|^{\mu}\right\rangle_{\text{QTM}}=\sum_{i=1}^{\mu} \frac{\Gamma\left[1+\gamma_{\mu}^{(i)}\right]}{A^{\gamma_{\mu}^{(i)}} \Gamma\left[1+\alpha \gamma_{\mu}^{(i)}\right]} \frac{B_{\mu}^{(i)}}{\Lambda^{\gamma_{\mu}^{(i)}}} t^{\alpha \gamma_{\mu}^{(i)}}.
\end{equation}
The constants $\gamma _{\mu}^{(i)}$, $B_{\mu}^{(i)}$ and $\Lambda$ (which is determined by $Q_0$ through Eq. (\ref{eq: Lambda theory})) depend on the lattice dimension, force applied and the type of lattice. In the next section, we explicitly evaluate the first and second moments of the biased QTM model for a few two-dimensional lattice types.

\section{Response to a bias} \label{se: Response to a bias}
In this section, we use the general form for the moments of $\mathbf{r}$ as a function of measurement time (Eq. (\ref{eq: final moments})) and evaluate the particular values of the first and second moments. We then utilize these quantities to explicitly obtain the response to bias in QTM, i.e., the case when an external force is applied. In this section, we consider the  rectangular and square lattice. Appendix (Sec. \ref{se: Appendix}) shows how our approach can be extended to other lattice types, such as the oblique and hexagonal.

To calculate the moments of the QTM in Eq. (\ref{eq: final moments}), we need to find the constants $\gamma _{\mu}^{(i)}$, $B_{\mu}^{(i)}$ ($i=1..\mu$) and $\Lambda$ (which is determined by the return probability to a previously visited site $Q_0$). These constants are determined by the dimension, lattice type, and properties of the applied force.
By applying a bias, i.e., a force $F_0$, we modify the constants $B_{\mu}^{(i)}$ and the return probability $Q_0$ (which determines $\Lambda$ in Eq. (\ref{eq: final moments})). 
First we evaluate $\gamma _{\mu}^{(i)}$ and $B_{\mu}^{(i)}$ ($i=1..\mu$) by calculating the moments of the spatial processes in Eq. (\ref{eq: moments spatial process}), i.e regular RW on top of a lattice.
We consider the case of a rectangular lattice with step size $a$ in the $x$ direction and $b$ in the $y$ direction.
By setting $a=b$, we can  easily find the case of the square lattice. We define $q_{\rightarrow}$ and $q_{\uparrow}$ to be the probability of the walker to take a step to the right ($x$ axis) and upwards ($y$ axis), respectively. The probabilities to jump to the left and downwards are then simply $q_{\leftarrow}=(1/2 - q_{\rightarrow})$ and $q_{\downarrow}=(1/2 - q_{\uparrow})$ respectively. 
So only two of the four jumping probabilities are required to define the problem. 
The external force $F_0$ is applied to the system at an angle $\theta$ (relative to the positive direction of the $x$ axis). 
By applying a force, we change the hopping probabilities of the RW \cite{PhysRevE.67.026128, PhysRevE.69.026103, J.Phys.Condens.31.445401}. 
When the applied external force is sufficiency small (and independent of time), the jumping probabilities are proportional to $\exp (aF_x/2k_B T)$ and $\exp (bF_y/2k_B T)$ for a jump in the $x$ direction and the $y$ direction, respectively. 
Here $k_B$ is Boltzmann's constant, $T$ is the temperature, $F_x$ and $F_y$ are the projections of the force on the $x$ and $y$ axis, i.e. $F_x=F_0 \cos (\theta)$ and $F_y=F_0 \sin (\theta)$. 
Since we are interested in the limit of small (and constant) force, the exponential contribution can be approximated to be linear by taking only the first two terms in Taylor's series expansion. Hence the jumping probabilities for a single step are
\begin{equation}
\begin{aligned}
    q_{\rightarrow}&=\frac{1}{4}(1+\widetilde{F}_x)\\ q_{\uparrow}&=\frac{1}{4}(1+\widetilde{F}_y)\\
    q_{\leftarrow}&=\frac{1}{4}(1-\widetilde{F}_x)\\
    q_{\downarrow}&=\frac{1}{4}(1-\widetilde{F}_y)
\end{aligned}
\end{equation}
where $\widetilde{F}_x=a F_0 \cos (\theta) / (2 k_B T)$ and $\widetilde{F}_y=b F_0 \sin (\theta) / (2 k_B T)$. 
Note that $\widetilde{F}_x$ and $\widetilde{F}_y$ are both dimensionless variables. By using the expressions for jumping probabilities, the moments of the horizontal position $x$ can be obtained for the RW process on top of a rectangular lattice. 
We label these moments as $\big\langle x \big\rangle_{RW}$ and $\big\langle x^2 \big\rangle_{RW}$ with subscript RW to distinguish them from the QTM's moments (which will be labeled with a subscript of QTM). 
The average value of the RW's position $\mathbf{r}$ after $N$ steps can be found from the average value of a single step displacement $l$ since $\big\langle \mathbf{r} \big\rangle_{RW} = \big\langle \sum_{i=1}^{N}l_i\big\rangle=N\langle l \rangle$ where $l_i$ is the walker's displacement at step number $i$ ($i\leq N$). 
The second moment is found by using the fact that the variance of $\mathbf{r}$ can be deconstructed into a sum of the variances for different steps. This is  true only when the steps are independent and uncorrelated. 
Hence the moments (say of $x$) of the spatial process are now
\begin{equation} \label{eq: first moment RW}
\big\langle x \big\rangle_{RW} = a(2q_{\rightarrow}-0.5)N
\end{equation}
\begin{equation} \label{eq: second moment RW}
\big\langle x^2 \big\rangle_{RW} =  a^2\left(0.5-(2q_{\rightarrow}-0.5)^2\right)N+a^2\left(2q_{\rightarrow}-0.5\right)^2 N^2
\end{equation}
Comparison of Eq. (\ref{eq: moments spatial process}) to Eqs. (\ref{eq: first moment RW},\ref{eq: second moment RW}) yields for $\gamma _{\mu}^{(i)}$ and $B_{\mu}^{(i)}$ ($i=1..\mu$)
\begin{equation} \label{eq: constants}
\begin{aligned} 
    B_1^{(1)}&=a(2q_{\rightarrow}-0.5), \;\; \gamma_1^{(1)}=1 \\
    B_2^{(1)}&=a^2(0.5-(2q_{\rightarrow}-0.5)^2), \;\; \gamma_2^{(1)}=1\\
    B_2^{(2)}&=a^2(2q_{\rightarrow}-0.5)^2, \;\; \gamma_2^{(2)}=2.
\end{aligned}
\end{equation}
The only missing component for computation of the moments for QTM is $Q_0$ that determines $\Lambda$  in Eq.~\eqref{eq: final moments} using Eq.~\eqref{eq: Lambda theory}. $Q_0$ depends on $F_0$ in a non-trivial fashion. Below we develop this dependence that is summarized in Eq.~\eqref{eq: Q_0 final}.

The return probability $Q_0$ is found by the means of generating functions~\cite{WeissBook}.
We use the probability of first return to the starting point ($\mathbf{r}=0$) after $N$ steps, i.e., $f_N (0)$, and represent $Q_0$ as  
$Q_0=\lim_{z\rightarrow1} \left( \sum_{N=0}^{\infty} f_N (0) z^N \right)$.
$p_N (0)$ is the probability to find the RW at position $\mathbf{r}=0$ after $N$ steps. 
The quantities $f_N (0)$ and $p_N(0)$ are related by~\cite{WeissBook}
\begin{equation}
    \sum_{N=0}^{\infty} f_N (0) z^N = 1-\frac{1}{\sum_{N=0}^{\infty} p_N (0) z^N}
\end{equation}
which means that
\begin{equation} \label{eq: Q_0 definition}
    Q_0= 1-\lim_{z\rightarrow1}\frac{1}{\left(\sum_{N=0}^{\infty} p_N (0) z^N\right)}.
\end{equation}
The term in the denominator in Eq. (\ref{eq: Q_0 definition}) is evaluated by calculating the Fourier transform of the jump probability of the RW, i.e., $\lambda(\mathbf{k})=\langle \exp (i \mathbf{k} \cdot \mathbf{r}) \rangle_{RW}$, and taking the inverse Fourier transform of the sum of geometrical series, i.e. 
\begin{equation} \label{eq:Izsum}
    I_z=\sum_{N=0}^{\infty} p_N (0) z^N=\frac{ab}{(2 \pi)^2} \int_{-\pi/b}^{\pi/b} \int_{-\pi/a}^{\pi/a} \frac{d^2\mathbf{k}}{1-z \lambda(\mathbf{k})}.
\end{equation}
In the case of rectangular lattice, $\lambda(\mathbf{k})$ is a finite sum
\begin{equation}
    \lambda(\mathbf{k}) = \langle e^{i\mathbf{k}\cdot\mathbf{r}}\rangle_{RW}=q_{\rightarrow} e^{i \widetilde{k}_x}+(1/2-q_{\rightarrow}) e^{-i \widetilde{k}_x}+q_{\uparrow} e^{i \widetilde{k}_y}+(1/2-q_{\uparrow}) e^{-i \widetilde{k}_y}
\end{equation}
where we set for convenience the change of variables $\widetilde{k}_x=a k_x$ and $\widetilde{k}_y=b k_y$.
We substitute $q_{\rightarrow}=\frac{1}{4}(1+\widetilde{F}_x)$ and $q_{\uparrow}=\frac{1}{4}(1+\widetilde{F}_y)$, that yields for Eq.~\eqref{eq:Izsum}

\begin{align} \label{eq:c_z2}
    I_z = \frac{1}{(2 \pi)^2}\int_{-\pi}^{\pi} \int_{-\pi}^{\pi} \frac{d\widetilde{k}_x d\widetilde{k}_y}{1-z\big[\frac{1}{2}\cos(\widetilde{k}_x)+\frac{1}{2}\cos(\widetilde{k}_y)+\frac{i}{2}\big(\widetilde{F}_x\sin(\widetilde{k}_x)+\widetilde{F}_y\sin(\widetilde{k}_y)\big)\big]}.
\end{align}
Note that when the external force $F_0>0$, this integral attains a final value in contrast to the unbiased case.
That is why we can substitute $z=1$ straight away before explicitly evaluating the integral. Additionally, multiplying both the nominator and denominator by the conjugate of the denominator yields
\begin{align}
    I_1 = \frac{1}{(2 \pi)^2}\int_{-\pi}^{\pi} \int_{-\pi}^{\pi} \frac{[1-\frac{1}{2}\cos(\widetilde{k}_x)-\frac{1}{2}\cos(\widetilde{k}_y)]+\frac{i}{2}[\widetilde{F}_x\sin(\widetilde{k}_x)+\widetilde{F}_y\sin(\widetilde{k}_y)]}{[1-\frac{1}{2}\cos(\widetilde{k}_x)-\frac{1}{2}\cos(\widetilde{k}_y)]^2+\frac{1}{4}[\widetilde{F}_x\sin(\widetilde{k}_x)+\widetilde{F}_y\sin(\widetilde{k}_y)]^2}d\widetilde{k}_x d\widetilde{k}_y
\end{align}
Since we are looking for a strictly real solution, we can take only the real part of the integral, i.e.,
\begin{align}
    I_1 = \frac{1}{(2 \pi)^2}\int_{-\pi}^{\pi} \int_{-\pi}^{\pi} \frac{\sin^2(\widetilde{k}_x/2)+\sin^2(\widetilde{k}_y/2)}{[\sin^2(\widetilde{k}_x/2)+\sin^2(\widetilde{k}_y/2)]^2+\frac{1}{4}[\widetilde{F}_x\sin(\widetilde{k}_x)+\widetilde{F}_y\sin(\widetilde{k}_y)]^2}d\widetilde{k}_x d\widetilde{k}_y
\end{align}
Both the nominator and denominator are strictly positive, and the main contribution to the integral is from the area when both $k_x, k_y \to 0$. We  approximate the solution by taking only the first terms in the Taylor series expansion of the trigonometric functions
\begin{align}
    I_1 = \frac{1}{(2 \pi)^2}\int_{-\pi}^{\pi} \int_{-\pi}^{\pi} \frac{4(\widetilde{k}_x)^2+4(\widetilde{k}_y)^2}{[(\widetilde{k}_x)^2+(\widetilde{k}_y)^2]^2+4[\widetilde{F}_x \widetilde{k}_x+\widetilde{F}_y \widetilde{k}_y]^2}d\widetilde{k}_x d\widetilde{k}_y.
\end{align}
By switching to polar coordinates $\rho=\sqrt{(\widetilde{k}_x)^2+(\widetilde{k}_y)^2}$ and $\phi=\arctan\left(\widetilde{k}_y / \widetilde{k}_x\right)$:
\begin{equation}
\begin{aligned}
    I_1 &= \frac{1}{(2 \pi)^2}\int_{0}^{2\pi} \int_{0}^{\pi\sqrt{2}} \frac{4\rho}{{\rho}^2+4\big(\widetilde{F}_x \cos \phi + \widetilde{F}_y \sin \phi\big)^2}d\rho d\phi\\
    &=\frac{2}{(2\pi)^2}\int_{0}^{2\pi} \left[ \ln\big[2\pi^2+4\big(\widetilde{F}_x \cos \phi + \widetilde{F}_y \sin \phi\big)^2\big] -\ln\big[4\big(\widetilde{F}_x \cos \phi + \widetilde{F}_y \sin \phi\big)^2\big] \right]d\phi.
\end{aligned}    
\end{equation}
and taking the limit $F_0 \rightarrow 0$ we obtain that
\begin{equation}
 I_1 \sim \frac{1}{\pi}\ln \left( \frac{2\pi^2}{(\widetilde{F}_x)^2+(\widetilde{F}_y)^2} \right). 
\end{equation}
Finally, plugging this form of $I_1$ into Eq. (\ref{eq: Q_0 definition}), we find the dependence of the return probability on the force,
\begin{equation} \label{eq: Q_0 final}
    Q_0=1+\pi \left[ \ln \left( \frac{(\widetilde{F}_x)^2+(\widetilde{F}_y)^2}{2\pi^2} \right)\right]^{-1} \text{  when  } F_0 \rightarrow 0.
\end{equation}
This result indicates that the return probability has a logarithmic dependence on $F_0$.

Now we can finalize the calculation of  the first and second moments for the QTM process. 
By plugging in the constants, $\gamma _{\mu}^{(i)}$, $B_{\mu}^{(i)}$ $(i=1..\mu)$ found from Eq. (\ref{eq: constants}) into Eq. (\ref{eq: final moments}) we obtain that for $0<\alpha<1$
\begin{equation} \label{eq: first moment theory}
    \big\langle x \big\rangle_{QTM} =
    \frac{a}{2 A \Gamma [1+\alpha]}\frac{\widetilde{F}_x}{\Lambda }t^{\alpha}
\end{equation}
and
\begin{equation} \label{eq: second moment theory}
    \big\langle x^2 \big\rangle_{QTM} 
    =
    \frac{a^2}{2A\Gamma[1+\alpha]}\left( 1-\frac{1}{2}(\widetilde{F}_x)^2 \right)\frac{1}{\Lambda }t^{\alpha}+\frac{a^2}{2A^2\Gamma[1+2\alpha]}\frac{(\widetilde{F}_x)^2}{\Lambda^2}t^{2\alpha}
\end{equation}
 while $\Lambda$ is provided by Eq.~\eqref{eq: Lambda theory} and $Q_0$ is the return probability given in Eq. (\ref{eq: Q_0 final}).
 
For the case of a square lattice, we set the step lengths in both the $x$ and $y$ direction equal, hence $a=b$. Therefore for a square lattice, the return probability is 
\begin{equation}
    Q_0=1+\frac{\pi}{2\ln \left( a\widetilde{F}_0 / (\sqrt{2} \pi) \right)} 
\end{equation}
where  $\widetilde{F}_0=\frac{F_0}{2k_BT}$. 
This implies that for a square lattice, the return probability has no dependence on the direction of the force. 

Our results show that the moments (Eq. (\ref{eq: first moment theory}-\ref{eq: second moment theory})) have non-linear response both in time ($\langle x \rangle \sim t^{\alpha}$) and in the applied force. In particular, besides anomalous diffusion, the  coefficients in both moments depend logarithmically on the external force applied through $\Lambda$. This is in contrast to CTRW, where the first and second moment has a coefficient that depends linearly on $F_0$ (in Sec.,\ref{se: Equivalency of the QTM to CTRW} we will see the exact connection between CTRW and the QTM). The moments of $x,y$ also show dependence on the direction through the force term ($\widetilde{F}_x$ or $\widetilde{F}_y$) and $Q_0$ (which determines $\Lambda$). This directional dependence will also appear in $\mathbf{r}$. In the case of the square lattice, the return probability ($Q_0$) does not depend on the direction of the force due to spatial symmetry, hence the moments of the displacement ($\mathbf{r}$) also do not dependent on the direction of the applied force.

Our results are consistent with the expectation of unbiased scenario when $F_0=0$. By using the asymptotic relation $\text{Li}_{-\alpha}(Q_0)\sim\Gamma[1+\alpha](1-Q_0)^{-\alpha-1}$~\cite{handbook.1972} in Eq.~\eqref{eq: first moment theory} the limit of the unbiased case, i.e., $\langle x \rangle_{QTM}=0$, is obtained in a straightforward manner. 
Note that the presented theory can be used to calculate the QTM's moments of any parameter ($x$, $y$ or $\mathbf{r}$) by switching only the relevant constants for $B_{\mu}^{(i)}$ and $\gamma_{\mu}^{(i)}$ in Eq. (\ref{eq: final moments}) as we will see in the appendix (Sec. \ref{se: Appendix}) for the cases of the oblique and hexagonal lattice types. 

\section{Simulations} \label{se: Simulations}
\begin{figure}[t]
\centering
\includegraphics[width=0.5\textwidth]{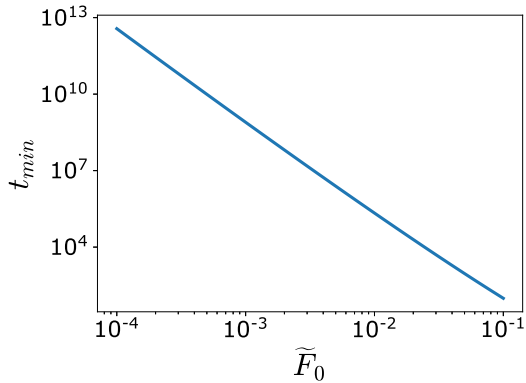}
\caption{The plot in the log-log scale of the minimal time ($t_{min}$) required for the theory to be a correct representation of the  simulations as described by Eq. (\ref{eq: minimal time}) vs. the force applied $\widetilde{F}_0$. The constants taken here are $\alpha=0.5$, $A=1$ and $\theta=\pi/4$. The lattice type is a square with step size $a=1$.}
\label{Fig: min time}
\end{figure}
\begin{figure}[t]
\centering
\includegraphics[width=0.5\textwidth]{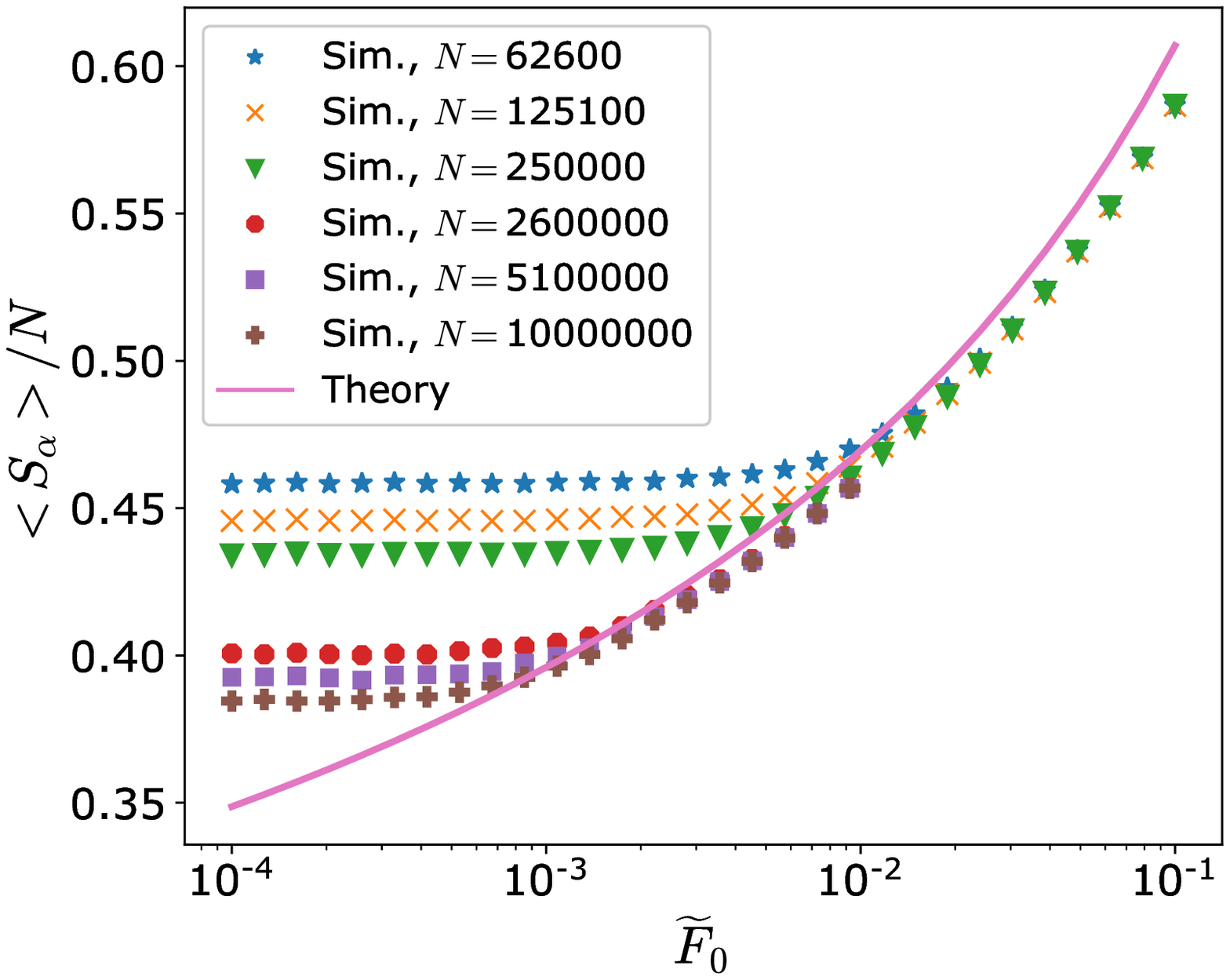}
\caption{Convergence of local time $S_\alpha$ to $\Lambda N$ on a two-dimensional square lattice. Plot of simulation results of $\Lambda = \langle S_{\alpha} (N) \rangle / N$ for increasing values of $N$ (number of steps) as a function of the force $\widetilde{F}_0$. The continuous line is the theoretical value of $\Lambda$ in Eq. (\ref{eq: Lambda theory}). The constants taken here are $\alpha=0.5$, $A=1$, $\theta=\pi/4$, $a=b=1$ and the ensemble average is done on $10^3$ independent RW's. Note that the difference between the theory and simulations in the range $\widetilde{F}_0>10^{-2}$ stems from the fact that we evaluated analytically $\Lambda$ (and its dependence on the return probability $Q_0$) in the limit $\widetilde{F}_0 \rightarrow 0$. Also note that as $N$ (number of steps) is increased we get a better agreement between theory and simulations for smaller values of $\widetilde{F}_0$. Taking larger values of N corresponds to larger values of measurement time $t$. The range of the plot corresponds to $Q_0\sim 0.853$ ($\widetilde{F}_0=10^{-4}$) and $\sim 0.586$ ($\widetilde{F}_0=10^{-1}$).}
\label{fig: Lambda}
\end{figure}
This section compares our analytical results to numerical simulations and discusses the numerical convergence to the theory. 

For the theory to correctly represent the simulations, we need to run the simulations for a sufficiently large measurement time $t$. 
This is because the 
theory was developed for the regime when $t\to\infty$ and $S_\alpha \to \infty$ such that a convergence of $S_\alpha$ to $\Lambda N$ (Eq. (\ref{eq: QTM local time linear connection})) occurs (as can be seen in Fig. \ref{fig: Lambda}).
This linear connection between $S_{\alpha}$ and $\Lambda$ is achieved  in the biased case when 
a sufficient number of returns of the RW to each site was performed. 
The RW must perform these returns; otherwise, the sites are not revisited, and the process will behave as CTRW, in contradiction to the theory. 
In other words, the correlations between various dwell times are essential for two-dimensional QTM. Therefore it is crucial to properly sample such events, i.e., observe enough returns of the RW.
Since there are more degrees of freedom in 2d than in 1d, the time it takes to sample such events in 2d is significantly larger than in 1d (as was already mentioned in the introduction).
Hence, for a given force $F_0$, it takes much more steps of the RW in 2d (compared to 1d) to observe the convergence of $S_\alpha$ to the linear prediction.
From these considerations, a logical approach is to  define, at least qualitatively,  minimum $t=t_{min}$, required for the theory (Eq. (\ref{eq: first moment theory}) and Eq. (\ref{eq: second moment theory})) to be correct.
Such a condition can be based on the physical intuition that the second moment $\langle x^2\rangle_{QTM}$ should always be monotonically increasing with the external force $F_0$.
Hence, in the case of a rectangular lattice  from the condition $\frac{\partial \langle x^2 \rangle}{ \partial \widetilde{F}_0} > 0$ we  find that for small values of $\widetilde{F}_0$
\begin{equation}
    t^{\alpha}>
    \frac{A\Gamma[1+2\alpha]}{2a^2\cos^2(\theta)\Gamma[1+\alpha]}\frac{1}{\widetilde{F}_0}\left(\frac{\partial \Lambda}{\partial \widetilde{F}_0}\right)
\end{equation}
where
\begin{equation}
    \frac{\partial \Lambda}{\partial \widetilde{F}_0}=
    \frac{2\pi\left(1-Q_0\right)}{\widetilde{F}_0\left( Q_0 \ln \left( \frac{a^2 \cos^2 (\theta)+b^2 \sin ^2 (\theta)}{2 \pi^2} (\widetilde{F}_0)^2 \right) \right)^2 } \left[ Li_{-\alpha} (Q_0) -(1-Q_0) Li_{-\alpha-1} (Q_0) \right].
\end{equation}
Hence the measurement time $t$ has to be larger then
\begin{equation} \label{eq: minimal time}
    t_{min}(\widetilde{F}_0)=\left( \frac{A\Gamma[1+2\alpha]}{2a^2\cos^2(\theta)\Gamma[1+\alpha]}\frac{1}{\widetilde{F}_0} \left( \frac{\partial \Lambda}{\partial \widetilde{F}_0} \right) \right)^{\frac{1}{\alpha}}
\end{equation}
for the theory to be correct for values of the external force as small as $\widetilde{F}_0$. We can notice that this value diverges when $\theta=\pi/2$. This is the case when the external force is applied in the $y$ direction, so the projection on the $x$ axis is zero.
In this case, the moments of $x$ do not depend on the force, and we need to apply the same logic to obtain the moments of $y$ instead. 
To see how $t_{min}$ behaves, we plot its dependence on the force In Fig. \ref{Fig: min time}. 
The asymptotic behavior of $t_{min}$ when $F_0\to 0$ is 
\begin{equation}
    t_{min} \sim \left(\widetilde{F}_0 \left| \ln(\widetilde{F}_0)\right|\right)^{-2/\alpha},
    \qquad \left(\widetilde{F}_0\to 0\right)
    \label{eq: asymptotic tmin}
\end{equation}
as can  also be verified in Fig. \ref{Fig: min time}. 
It is also apparent from Eq. (\ref{eq: minimal time}) that $t_{min}$ decreases as $\alpha$ is closer to $1$. 
This occurs since as $\alpha \rightarrow 1$ the system  approaches the conditions when mean dwell times are finite.
The time it takes $S_{\alpha}$ to converge to $\Lambda N$ is displayed in Fig. \ref{fig: Lambda}. We can see that as the force ($\widetilde{F}_0$) gets smaller, we need a larger amount of steps ($N$) for the convergence to take place. Taking a larger amount of steps corresponds to larger values of the measurement time ($t$). This can be seen from the PDF of $t$ given in Eq. (\ref{eq:t quenched relation}).
Note that the difference in Fig. (\ref{fig: Lambda}) between the theory and simulations in the range $\widetilde{F}_0>10^{-2}$ stems from the fact that we evaluated  $\Lambda$ (and its dependence on the return probability $Q_0$) in the limit $\widetilde{F}_0 \rightarrow 0$.

\begin{figure}[t]
\centering
\includegraphics[width=0.5\textwidth]{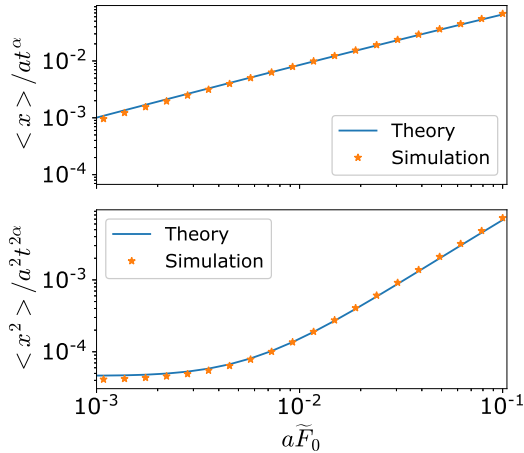}
\caption{A comparison between our theory and simulation results for the 2d QTM. In the top panel, we have a plot of the first moment and in the bottom the second moment. The lattice type taken here is a square with unit spacing ($a=1$). The constants are $\alpha =0.5$, $A=1$, $t=10^9$ and $\theta = \pi /4$. For each plot, the ensemble average is taken over $10^6$ samples. For the chosen value of $t$ ($t=10^9$) we have an agreement with theory for values of $\widetilde{F}_0$ as small as $\sim  10^{-3}$ (as explained in Sec. (\ref{se: Simulations}) (Simulations)).}
\label{Fig: moments}
\end{figure}

In Fig.~\ref{Fig: moments} we compare theory and simulation results of the first and second moments of the QTM.
For the value of measurement time $t$ chosen in Fig.~\ref{Fig: moments} ($t=10^9$) the theory and simulations agree for values of $\widetilde{F}_0$ as small as $\sim 3 \cdot 10^{-3}$. These forces and the measurement time (in Fig.~\ref{Fig: moments}) are consistent with Eq.~\eqref{eq: asymptotic tmin} . 

\section{Equivalency between the QTM and CTRW} \label{se: Equivalency of the QTM to CTRW}
\begin{figure}[t]
\centering
\includegraphics[width=0.5\textwidth]{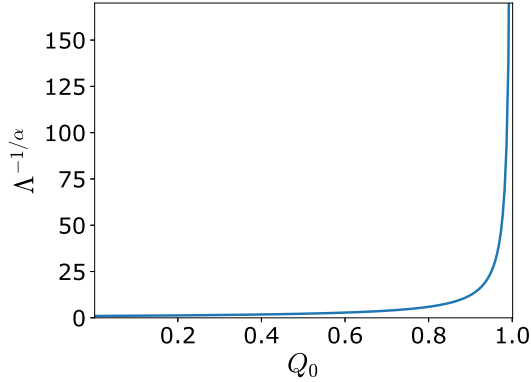}
\caption{A plot of the temporal transformation constant $1/ \Lambda ^{1/ \alpha}$ that takes us from CTRW to the equivalent state in QTM for $\alpha=0.5$ as a function of the return probability $Q_0$. The divergence for $Q_0\rightarrow1$ signifies the limitation of this transformation strictly to the transient case. Note that $\Lambda = 1$ when $Q_0 \rightarrow 0$, i.e when the force $F_0$ is large enough.}
\label{fig:3}
\end{figure}

In CTRW we have only one level of subordination since the local time is simply the step number $N$ as compared to $S_{\alpha}$ in QTM. Then since the PDF of CTRW (Eq. (\ref{eq: PDF CTRW})) is $\langle P(\mathbf{r}, t)\rangle_{CTRW}=\sum_{N} W_{N}(\mathbf{r}) \mathscr{P}_{t}(N)$ and $\mathscr{P}_{t}(N)$, the probability to observe $N$ steps during time $t$ is given in Eq. (\ref{eq: Pt(N)}) we have
\begin{equation} \label{eq: probability final CTRW}
\langle P(\mathbf{r}, t)\rangle_{CTRW} \sim \int_{0}^{\infty} W_{N}(\mathbf{r}) \frac{t}{\alpha} N^{-1 / \alpha-1} l_{\alpha, A, 1}\left(\frac{t}{N^{1 / \alpha}}\right) \mathrm{d} N \quad t \rightarrow \infty.
\end{equation}
Comparing this equation of CTRW to that of QTM (Eq. (\ref{eq: probability final})) yields the result
\begin{equation}
\langle P(\mathbf{r}, t)\rangle_{QTM} \sim \langle P(\mathbf{r}, t / \Lambda^{1 / \alpha})\rangle_{CTRW}, \; t\rightarrow\infty
\end{equation}
meaning that the transformation
\begin{equation}
    t \rightarrow t / \Lambda^{1 / \alpha}
\end{equation}
takes us from transient CTRW to the equivalent disorder averaged propagator in QTM. A plot of the constant ($1 / \Lambda^{1 / \alpha}$) is presented in Fig. \ref{fig:3}. When $Q_0=0$ ($\Lambda=1$), the QTM is exactly described by the CTRW since the walker never returns to a previously visited site. For any $0<Q_0<1$ the constant $(\Lambda^{-1/\alpha})$ is greater then $1$. This means that the limiting PDF attained in both models as $t\rightarrow \infty$ is achieved faster in the QTM than in the CTRW. The reason is that CTRW has a higher chance (compared to QTM) to roll a very large dwell time at each step (because we re-roll a new dwell time every step regardless of the position in space) and thus "waste" more time.

Furthermore, we compare this temporal mapping to the equivalent one in the one-dimensional case. In the case of the biased $1$d QTM, we know from \cite{JStatMech2020.073207} that the escape probability is
\begin{equation}
    \widetilde{Q}_0=2(1-q)
\end{equation}
where $q$ is the bias in $1$d to step right.
From our definition of the force in $2$d the applied force on the $x$ axis is $\widetilde{F}_x= a \cos (\theta) \widetilde{F}_0$. By defining similarly the connection between the force and bias in $1$d, we will obtain $q=\frac{1}{2}(1+\widetilde{F}_x)$, i.e, $\widetilde{Q}_0=1- \widetilde{F}_x$. While in $2$d, say for the square lattice, we obtained in Eq. (\ref{eq: Q_0 definition}) that the return probability is $Q_0=1+0.5\pi / \ln \left( a\widetilde{F}_0 / (\sqrt{2} \pi) \right)$.
The comparison of the transformation constant $1/\Lambda^{1/\alpha}=\left[ Q_0 / \big( (1-Q_0)^2 Li_{-\alpha}(Q_0) \big)\right]^{1/\alpha}$ (from Eq. (\ref{eq: Lambda theory})) for the $1$d case against the $2$d square lattice case 
shows that for a given force $\widetilde{F}_0$, $1/\Lambda^{1/\alpha}$ is larger for the $1$d case. 
Meaning that the attained limiting PDF, when $t\rightarrow\infty$, is reached faster for the QTM in $1$d than in $2$d when the same force $\widetilde{F}_0$ is applied.
This can be explained by the fact that in $2$d, we have more degrees of freedom and thus, on average, during time $t$, the RW visits a larger amount of different sites and, in turn, have more chances to roll larger dwell times and "waste" more time. 
The re-visitation rate for $1$d is higher than in $2$d. 
It also answers why the $2$d QTM is closer to its CTRW counterpart than to the $1$d version.

\section{Conclusions} \label{se: Conclusions}
By using the method of double subordination and utilizing the local time $S_{\alpha}$,   we obtain the positional PDF for the two-dimensional biased QTM with external force $F_0$. This PDF and the obtained return probability $Q_0$ allows us to write an explicit formula for the moments of the system on various lattice types in $2$d. 
The moments show non-linear response both in time ($\langle x \rangle \sim t^{\alpha}$) and in the applied force. 
Our results are consistent with Ref., \cite{PhysRevE.101.042133} which predicts non-self-averaging, characteristics of aging, and anomalous behavior when $\alpha<1$, i.e., diverging mean dwell times. 
In particular, besides anomalous behavior in time, our results show that the  first and second moments depend logarithmically on the external force applied via the $\Lambda$ term. 
This result is in contrast to the CTRW model, where the first moment has a coefficient that depends linearly on $F_0$.
The moments are also affected in a non-linear fashion by the direction of the applied force in each lattice type that was presented (rectangular, oblique, and hexagonal) except for the square lattice type, where the dependence is canceled out due to spatial symmetry. 
This again is in contrast to what is expected for the mean-field representation of CTRW \cite{Klafter.Sokolov.Book}. These differences between QTM and CTRW appear despite the equivalence between the two models (when external bias is present) that is achieved via the temporal transformation $t\to t/\Lambda^{1/\alpha}$. The constant $\Lambda$ depends on the direction and the size of the force ${\widetilde F}_0$ in a non-linear fashion.  

Quenched disorder and anomalous behavior of the dwell times induce the observed non-linear response, i.e., non-linear dependence of the average position on the size of the external field. 
The non-linear response and associated breaking of the Einstein relation were predicted for the $1$d QTM using scaling arguments~\cite{PhysRep.195.127}, and explicitly calculated in the limit of $T\to 0$~\cite{PhysRevE.68.036114}. This non-linear response appears in $1$d only for measurement times $t>({\widetilde F}_0)^{-1-1/\alpha}$ when ${\widetilde F}_0\to 0$~\cite{PhysRep.195.127}.  
Our results (Eq.~\eqref{eq: asymptotic tmin}) show that for the two-dimensional case the measurement time $t$ for which the non-linear response reveals itself scales as  $\left(\widetilde{F}_0 \left| \ln(\widetilde{F}_0)\right|\right)^{-2/\alpha}$. For any $0<\alpha<1$ the time-scales where non-linearity of response takes place is therefore significantly larger for the two-dimensional case.
Since the QTM was developed as a toy model for glassy systems, this prediction of non-linear response to external force in $2$d might be explored in situations where aging, quenched disorder, and anomalous behavior have been reported. Systems like Physical glasses~\cite{AmirPnas,BethierBiroli} and bio-materials~\cite{KrapfPRX,SlezakBurov}. It is important to remember that our work suggests that this non-linearity will reveal itself only for a very long measurement time.    
The last thing to mention is that our theory can be extended to other lattices in higher dimensions as long as the return probability $Q_0$ is known and smaller than one.

\section*{Acknowledgements}
This work was supported by the Israel Science Foundation Grant No. 2796/20. 

\section{Appendix} \label{se: Appendix}

\subsection{Appendix A - Oblique lattice} \label{subsection: oblique and hexagonal}
\begin{figure}[h]
\centering
\includegraphics[width=0.5\textwidth]{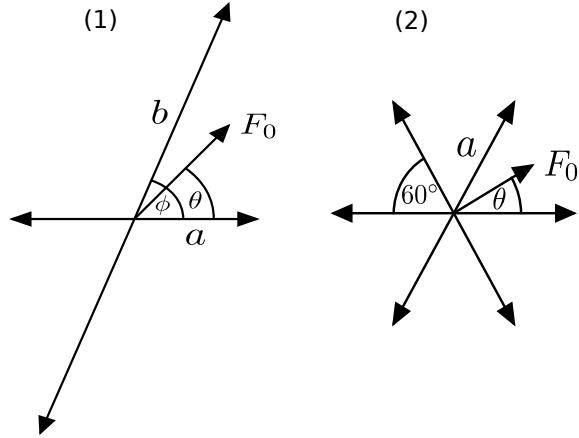}
\caption{A sketch of the two lattice types examples is discussed in this section. In panel (1), we have a general oblique lattice skewed by an angle of $\phi$ relative to the $x$ axis. In panel (2), a hexagonal lattice type is sketched, the walker has at each new random step six evenly spaced out directions to choose from.}
\label{fig: lattice drawing}
\end{figure}

For the oblique lattice, we have a walker that can jump a distance of $a$ in either direction on the $x$ axis and a distance of $b$ in either direction on the axis rotated by an angle of $\phi$ (relative to the $x$ axis). An example of a general oblique lattice can be seen in Fig. \ref{fig: lattice drawing} panel (1). Following the same logic as in the rectangular lattice type, the jumping probabilities for a single step on an oblique lattice are
\begin{equation}
\begin{aligned}
    q_{a}&=\frac{1}{4}(1+\widetilde{F}_a)\\ q_{b}&=\frac{1}{4}(1+\widetilde{F}_b)\\
    q_{-a}&=\frac{1}{4}(1-\widetilde{F}_a)\\
    q_{-b}&=\frac{1}{4}(1-\widetilde{F}_b)
\end{aligned}
\end{equation}
where from geometrical considerations, we have the bias in each direction
\begin{equation}
\begin{aligned}
\widetilde{F}_a&=\frac{aF_0}{2k_B T}\frac{\sin (\phi-\theta)}{\sin (\phi)}\\
\widetilde{F}_b&=\frac{bF_0}{2k_B T}\frac{\sin (\theta)}{\sin (\phi)}.
\end{aligned}
\end{equation}
$\widetilde{F}_a$ is the bias in the direction of $x$ and $\widetilde{F}_b$ is the bias in the direction of the step size $b$, i.e., in the direction of $\phi$ relative to the positive direction of $x$ (see Fig. \ref{fig: lattice drawing}). 

We repeat the same steps as in the rectangular lattice to calculate the return probability $Q_0$ of the oblique lattice.
The characteristic function $\lambda(\mathbf{k})=\langle \exp (i \mathbf{k} \cdot \mathbf{r}) \rangle$ in the case of the oblique lattice becomes 
\begin{equation}
\begin{aligned}
    \lambda(\mathbf{k})=&\frac{1}{4}(1+F_a)\exp (iak_x )+\frac{1}{4}(1-F_a)\exp (-iak_x)\\+&\frac{1}{4}(1+F_b)\exp (ib \cos (\phi) k_x  +ib \sin (\phi) k_y )\\+&\frac{1}{4}(1-F_b)\exp (-ik_x b \cos (\phi)-i k_y b \sin (\phi)).
\end{aligned}
\end{equation}
Now transforming $ \lambda(\mathbf{k})$ to trigonometric functions we receive
\begin{equation}
\begin{aligned}
    \lambda(\mathbf{k})=&\frac{1}{2}\cos(ak_x)+\frac{1}{4}iF_a\sin(ak_x)\\+&\frac{1}{2}\cos(ib \cos (\phi) k_x  +ib \sin (\phi) k_y)+\frac{1}{2}\cos(ib \cos (\phi) k_x  +ib \sin (\phi) k_y).
\end{aligned}
\end{equation}
Our boundaries in this case are $|x|<\text{max} \{ a,b\cos (\phi) \} $ since the projection of $b$ onto the $x$ axis can be larger then the step size $a$ on the $x$ axis and $|y|<b \sin(\phi)$. The integral we need to evaluate in Eq. (\ref{eq:Izsum}) is now 
\begin{equation} \label{eq:Izsum oblique}
    I_z=\frac{c b \sin (\phi)}{4 \pi^2} \int_{-\frac{\pi}{b \sin (\phi)}}^{\frac{\pi}{b \sin (\phi)}} \int_{-\pi/c}^{\pi/c} \frac{dk_x dk_y}{1-z \lambda(k_x, k_y)}
\end{equation}
where we set $c=\text{max} \{ a,b\cos (\phi) \}$. By using Taylor expansion in the $k_x,k_y\rightarrow0$ limit and substituting $z=1$, Eq. (\ref{eq:Izsum oblique}) becomes
\begin{equation}
    I_1=\frac{c b \sin (\phi)}{4 \pi^2} \int_{-\frac{\pi}{b \sin (\phi)}}^{\frac{\pi}{b \sin (\phi)}} \int_{-\pi/c}^{\pi/c} 
    \frac{dk_x dk_y \left[ {(ak_x/2)}^2 +{(\Vec{k} \cdot \Vec{b}/2)}^2 \right]}{\left[ {(ak_x/2)}^2 +{(\Vec{k} \cdot \Vec{b}/2)}^2 \right]^2+\left( a\widetilde{F}_a k_x /2 + \widetilde{F}_b \Vec{k} \cdot \Vec{b} /2 \right)^2}.
\end{equation}
Changing variables $\widetilde{k}_x=ak_x$ and $\widetilde{k}_y=k_x b \cos (\phi) +k_y b \sin(\phi)$ the Jacobian becomes $J=1/(ab\sin(\phi))$ and $I_1$ is transformed to
\begin{align}
    I_1 = \frac{c}{4 \pi^2 a}\int_{-\pi-\pi b \cos(\phi) /c}^{\pi+\pi b \cos(\phi) /c} \int_{-a\pi/c}^{a\pi/c} \frac{4(\widetilde{k}_x)^2+4(\widetilde{k}_y)^2}{[(\widetilde{k}_x)^2+(\widetilde{k}_y)^2]^2+4[\widetilde{F}_x \widetilde{k}_x+\widetilde{F}_y \widetilde{k}_y]^2}d\widetilde{k}_x d\widetilde{k}_y.
\end{align}
By switching to polar coordinates and taking the limit $F_0\rightarrow0$, we obtain 
\begin{align}
    I_1=\frac{c}{a\pi} \ln \left( \frac{\left( \frac{a\pi}{c}\right)^2+\pi^2\left( 1+\frac{b\cos(\phi)}{c} \right)^2}{(\widetilde{F}_a)^2+(\widetilde{F}_b)^2} \right)
\end{align}
and since $Q_0=1-1/I_1$ the final form for the return probability in the $F_0\rightarrow0$ limit is
\begin{equation}
    Q_0=1+\frac{a\pi}{c}\left[ \ln \left( \frac{({\widetilde{F}_a})^2+({\widetilde{F}_b})^2}{\left( \frac{a\pi}{c}\right)^2+\pi^2\left( 1+\frac{b\cos(\phi)}{c} \right)^2} \right)\right]^{-1}
    \qquad (F_0\rightarrow0).
\end{equation}
Now that we know the return probability $Q_0$, we have the value of $\Lambda$ from Eq. (\ref{eq: Lambda theory}). All we require now to have the value of the moments are the constants of the spatial processes $\gamma _{\mu}^{(i)}$, $B_{\mu}^{(i)}$ ($i=1..\mu$) for the case of the oblique lattice. After we find these constants, we plug them into Eq. (\ref{eq: final moments}) and acquire the moments.

We now continue to find the constants of the spatial processes. The average displacement after $N$ steps is related to a single step displacement $l_x$ by $\langle x \rangle = N \langle l_x \rangle$, from here we find
\begin{equation} \label{eq: first moment RW oblique}
\langle x \rangle_{RW} = (2a\widetilde{F}_a+2\cos (\phi) b \widetilde{F}_b)N.
\end{equation}
And since the steps are independent of each other (Markovian process) the variance of the position $x$ after $N$ steps is related to the variance of a single step displacement $l_x$ by $\text{Var}(x)=N\text{Var}(l_x )$.
The second moment of the spatial process now becomes
\begin{equation} \label{eq: second moment RW oblique}
\begin{aligned}
    \langle x^2 \rangle_{RW} =& \left[ \left(\frac{1}{2}a^2+\frac{1}{2}b^2\cos^2(\phi)\right)-\left(2a\widetilde{F}_a+2\cos (\phi) b \widetilde{F}_b\right)^2\right]N\\ &+\left[2a\widetilde{F}_a+2\cos (\phi) b \widetilde{F}_b\right]^2 N^2
\end{aligned}
\end{equation}
Comparison of Eq. (\ref{eq: moments spatial process}) to Eqs. (\ref{eq: first moment RW oblique},\ref{eq: second moment RW oblique}) yields for $\gamma _{\mu}^{(i)}$ and $B_{\mu}^{(i)}$ ($i=1..\mu$)
\begin{equation} \label{eq: constants oblique}
\begin{aligned} 
    B_1^{(1)}&=2a\widetilde{F}_a+2\cos (\phi) b \widetilde{F}_b, \;\; \gamma_1^{(1)}=1 \\
    B_2^{(1)}&= \left(\frac{1}{2}a^2+\frac{1}{2}b^2\cos^2(\phi)\right)-\left(2a\widetilde{F}_a+2\cos (\phi) b \widetilde{F}_b\right)^2, \;\; \gamma_2^{(1)}=1\\
    B_2^{(2)}&=(2a\widetilde{F}_a+2\cos (\phi) b \widetilde{F}_b)^2, \;\; \gamma_2^{(2)}=2.
\end{aligned}
\end{equation}
This concludes the case of the oblique lattice.
\subsection{Appendix B - Hexagonal lattice}
We perform the same steps as in the rectangular lattice to explicitly calculate the return probability for the hexagonal lattice. The hexagonal lattice has six symmetrical jump directions of size $a$ in each step (as sketched in Fig. \ref{fig: lattice drawing} panel (2)), the characteristic function $\lambda(\mathbf{k})=\langle \exp (i \mathbf{k} \cdot \mathbf{r}) \rangle$ becomes 
\begin{equation}
\begin{aligned}
    \lambda(\mathbf{k})&=\frac{1}{6}(1+\widetilde{F}_a)\exp(i a k_x)+\frac{1}{6}(1-\widetilde{F}_a)\exp(-i a k_x)+\frac{1}{6}(1+\widetilde{F}_b)\exp(i \frac{a}{2} k_x+i \frac{\sqrt{3} a}{2} k_y)\\
    &+\frac{1}{6}(1-\widetilde{F}_b)\exp(-i \frac{a}{2} k_x-i \frac{\sqrt{3} a}{2} k_y)+\frac{1}{6}\exp(-i\frac{a}{2}k_x+i\frac{\sqrt{3}a}{2}k_y)\\
    &+\frac{1}{6}\exp(i\frac{a}{2}k_x-i\frac{\sqrt{3}a}{2}k_y)
\end{aligned}
\end{equation}
where from geometrical considerations, we have
\begin{equation}
\begin{aligned}
    \widetilde{F}_a&=\left( \cos (\theta) -\frac{1}{\sqrt{3}}\sin (\theta)\right)\frac{aF_0}{2k_B T}\\
    \widetilde{F}_b&=\frac{aF_0 \sin (\theta)}{\sqrt{3}k_B T}.
\end{aligned}
\end{equation}
$\widetilde{F}_a$ is the bias in the direction of $x$ and $\widetilde{F}_b$ is in the direction of $60^{\circ}$ relative to the positive direction of $x$ (see Fig. \ref{fig: lattice drawing}).
Now transforming $\lambda(\mathbf{k})$ to trigonometric functions we receive
\begin{equation}
\begin{aligned}
    \lambda(\mathbf{k})&=\frac{1}{3}\cos (a k_x)+\frac{1}{3}i\widetilde{F}_a \sin(a k_x)+\frac{1}{3}\cos (\frac{1}{2}a k_x+\frac{\sqrt{3}}{2}a k_y)\\
    &+\frac{1}{3}i\widetilde{F}_b\sin (\frac{1}{2}a k_x+\frac{\sqrt{3}}{2}a k_y)+\frac{1}{3}\cos (\frac{1}{2}a k_x-\frac{\sqrt{3}}{2}a k_y).
\end{aligned}
\end{equation}
Since our boundaries are $-a<x<a$ and $\frac{-\sqrt{3}a}{2}<y<\frac{\sqrt{3}a}{2}$ the integral we need to evaluate is
\begin{equation} \label{eq:Izsum hexagonal}
    I_z=\frac{\sqrt{3}a^2}{8 \pi^2} \int_{\frac{-2\pi}{\sqrt{3}a}}^{\frac{2\pi}{\sqrt{3}a}} \int_{-\pi/a}^{\pi/a} \frac{dk_x dk_y}{1-z \lambda(k_x, k_y)}.
\end{equation}

By using Taylor expansion in the $k_x,k_y\to 0$ limit, substituting $z=1$ and setting $\widetilde{k}_x=ak_x, \; \widetilde{k}_y=ak_y$, Eq.~\eqref{eq:Izsum hexagonal} is transformed into
\begin{equation}
    I_1=\frac{\sqrt{3}}{8 \pi^2} \int_{\frac{-2\pi}{\sqrt{3}}}^{\frac{2\pi}{\sqrt{3}}} \int_{-\pi}^{\pi}\frac{4 d\widetilde{k}_x d\widetilde{k}_y \left( {\widetilde{k}_x}^2 +{\widetilde{k}_y}^2 \right)}{\left( {\widetilde{k}_x}^2 +{\widetilde{k}_y}^2 \right)^2+\frac{64}{36}\left( \left( \widetilde{F}_a+\frac{1}{2}\widetilde{F}_b \right)\widetilde{k}_x+\frac{\sqrt{3}}{2}\widetilde{F}_b\widetilde{k}_y \right)^2}.
\end{equation}
By switching to polar coordinates 
and taking the limit $F_0 \rightarrow 0$ we obtain
\begin{equation}
    I_1=\frac{\sqrt{3}}{2\pi}\ln \left[ \frac{21\pi^2/4}{({\widetilde{F}_a})^2+({\widetilde{F}_b})^2+\widetilde{F}_a \widetilde{F}_b} \right].
\end{equation}
Since $Q_0=1-1/I_1$ the final form for the return probability in the $F_0\to 0$ limit is:
\begin{equation}
    Q_0=1+\frac{2\pi}{\sqrt{3}}\left[ \ln \left( \frac{({\widetilde{F}_a})^2+({\widetilde{F}_b})^2+\widetilde{F}_a \widetilde{F}_b}{21 \pi^2 /4} \right)\right]^{-1}
    \qquad (F_0\rightarrow0).
\end{equation}

To obtain the the moments of QTM on a hexagonal lattice,  constants of the spatial processes $\gamma _{\mu}^{(i)}$, $B_{\mu}^{(i)}$ ($i=1..\mu$) are needed. 
Following the same logic as before, we receive
\begin{equation} \label{eq: first moment RW hexagonal}
\langle x \rangle_{RW} = (2a\widetilde{F}_a+ a\widetilde{F}_b)N
\end{equation}
and
\begin{equation} \label{eq: second moment RW hexagonal}
    \langle x^2 \rangle_{RW} = \left[ \frac{1}{2}a^2-(2a\widetilde{F}_a+a\widetilde{F}_b)^2\right]N +(2a\widetilde{F}_a+a\widetilde{F}_b)^2 N^2.
\end{equation}
Comparison of Eq. (\ref{eq: moments spatial process}) to Eqs. (\ref{eq: first moment RW hexagonal},\ref{eq: second moment RW hexagonal}) yields for $\gamma _{\mu}^{(i)}$ and $B_{\mu}^{(i)}$ ($i=1..\mu$)
\begin{equation} \label{eq: constants hexagonal}
\begin{aligned} 
    B_1^{(1)}&=2a\widetilde{F}_a+ a\widetilde{F}_b, \;\; \gamma_1^{(1)}=1 \\
    B_2^{(1)}&=\frac{1}{2}a^2-(2a\widetilde{F}_a+a\widetilde{F}_b)^2, \;\; \gamma_2^{(1)}=1\\
    B_2^{(2)}&=(2a\widetilde{F}_a+a\widetilde{F}_b)^2, \;\; \gamma_2^{(2)}=2.
\end{aligned}
\end{equation}
This concludes the case of the hexagonal lattice.

\end{document}